\documentclass[preprint,prb,nopacs,superscriptaddress]{revtex4}
\usepackage{epsfig}
\begin{document}
\topmargin-1.0cm

\title {
Temperature dependent optical studies of Ti$_{1-x}$Co$_x$O$_2$}
\author{S. Guha}\email[Corresponding author E-mail:]{sug100f@smsu.edu}
\affiliation {Department of Physics, Astronomy and Materials
Science, Southwest Missouri State University, Springfield, MO
65804 USA}
\author {K. Ghosh} \author {J.G. Keeth}
\affiliation {Department of Physics, Astronomy and Materials
Science, Southwest Missouri State University, Springfield, MO
65804 USA}
\author {S. B. Ogale}
\author {S.R. Shinde}
\affiliation {Center for Superconductivity Research, Department of
Physics, University of Maryland, College Park, MD 20742-4111}
\author{J. R. Simpson} \author{H. D. Drew} \author{T. Venkatesan}
\affiliation {Department of Physics, University of Maryland,
College Park, MD 20742-4111}

\date{\today}

\begin{abstract}
We present the results of Raman and photoluminescence (PL) studies
on epitaxial anatase phase Ti$_{1-x}$Co$_x$O$_2$  films for $x$ =
0-0.07, grown by pulsed laser deposition. The low doped system
($x$=0.01 and 0.02) shows a Curie temperature of ~700 K in the
as-grown state. The Raman spectra from the doped and undoped films
confirm their anatase phase. The photoluminescence spectrum is
characterized by a broad emission from self-trapped excitons (STE)
at 2.3 eV at temperatures below 120 K. This peak is characteristic
of the anatase-phase TiO$_2$ and shows a small blueshift with
increasing doping concentration. In addition to the emission from
STE, the Co-doped samples show two emission lines at 2.77 eV and
2.94 eV that are absent in the undoped film indicative of a
spin-flip energy.

\end{abstract}

 \maketitle

Dilute magnetic semiconductors (DMS) based on II-VI
semiconductors, such as Mn-doped CdTe and ZnSe, \cite{furdyna} and
III-Vs, such as Mn-doped GaAs \cite{ohno} have been studied
extensively in the last decade due to their potential usage of
both charge and spin degrees of freedom of carriers in electronic
devices. However, these materials undergo a ferromagnetic phase
transition temperature much below room temperature. The discovery
of ferromagnetism with Curie temperatures above 300 K in
cobalt-doped TiO$_2$ has generated considerable amount of interest
in this system and similar oxide based ferromagnetic
semiconductors.\cite{matsumoto,shinde} First-Principles
calculations of Co-doped TiO$_2$ elucidate on the electronic
structure and microscopic distribution of Co atoms crucial for
ferromagnetic ordering.\cite{park,yang} Recently, Mn-doped GaN
films are also known to show room temperature
ferromagnetism.\cite{Reedgan,sasakigan}

TiO$_2$ is a wide band gap semiconductor with excellent
transmittance in the visible and near-infrared regions, high
refractive index, and high dielectric constant. It occurs in
three crystal structures: rutile, anatase, and brooklite. Anatase
TiO$_2$ is composed of stacked edge-sharing octahedrons formed by
six oxygen anions. There are significant differences in the
electronic structure between the anatase and rutile crystal
structures. Anatase TiO$_2$ has a very shallow donor level and
high electron mobilities compared to the rutile phase.\cite{forro}
The $n$-type carriers in TiO$_2$ result from oxygen vacancies,
which provide tunability to its properties. Detailed
investigations of the absorption edges in anatase and rutile
phases by Tang \textit{et al.} have shown that that the excitonic
states of the anatase phase are self-trapped while that of the
rutile phase are free.\cite {tang1}

In this work we present the results of Raman and photoluminescence
(PL) studies on epitaxial anatase phase Ti$_{1-x}$Co$_x$O$_2$
films for $x$ = 0, 0.01, 0.02, 0.04, and 0.07, grown by pulsed
laser deposition (PLD). Thin films of Ti$_{1-x}$Co$_x$O$_2$ were
deposited at a pressure of 2-4$\times$10$^{-5}$ on LaAlO$_3$ (001)
substrates by PLD, using sintered targets synthesized by the
standard solid-state route. The substrate temperature, laser
energy, and pulse repetition rate were kept at 700 $\rm ^o$C, 1.8
J/cm$^2$, and 10 Hz, respectively. The thickness of the films were
$\approx$100nm.

X-ray diffraction (XRD) was employed for structural
characterization. The hysteresis loops and magnetization were
obtained using a SQUID magnetometer. The Raman measurements were
carried out in a backscattering geometry using a fiber-optically
coupled confocal micro-Raman system (TRIAX 320) equipped with a
liquid N$_2$ cooled charge coupled detector. The 514.5 nm line of
an $\rm Ar^+$ ion laser was the excitation source. The microscope
is equipped with a holographic super-notch filter to block the
elastically scattered light. A 50 X microscope objective was used
to focus and collect the scattered laser light, with a spatial
resolution of about 5$\mu$m. The PL spectra were excited using the
363.8 nm line of an $\rm Ar^+$ laser. The luminescence excitation
was analyzed with a SPEX 0.85 m double monochromator equipped with
a cooled GaAs photomultiplier tube and standard photon counting
electronics. For low temperature measurements a closed cycle
refrigerator was employed.

\begin{figure}
\unitlength1cm
\begin{picture}(5.0,8.0)
\put(-2.5,-0.7){ \epsfig{file=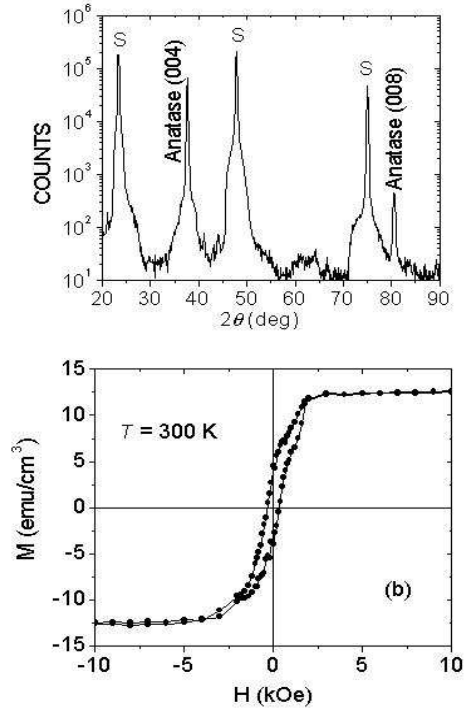, angle=0, width=9cm,
totalheight=11cm}}

\end{picture}
\caption{(a) XRD pattern and (b) 300 K hysteresis loop of
Ti$_{0.98}$Co$_{0.02}$O$_2$. In (a) the substrate peaks are
labelled by "s".} \label{fig1}
\end{figure}

Figure \ref{fig1} (a) and (b) shows the XRD pattern and the
magnetic response, respectively, of Ti$_{0.98}$Co$_{0.02}$O$_2$ on
LaAlO$_3$ (001). The (004) and the (008) peaks of the anatase
phase are clearly observed in the XRD. The XRD rocking curve full
width at half maximum of the sample peaks are $\sim$0.3 $\rm ^o$.
The magnetic response was measured as a function of the magnetic
field strength at room temperature. Hysteresis is observed
indicating that the film is ferromagnetic even at room
temperature. The low doped system ($x$=0.01 and 0.02) shows a
Curie temperature of $\sim$ 700 K in the as-grown
state.\cite{shinde}
The anatase and the rutile phase belong to the symmetry space
groups D$^{19}_{4h}$ and D$^{14}_{4h}$, respectively.\cite{ocana}
Group theory predicts six Raman active modes for the anatase
phase; E$_g$: 147 cm$^{-1}$, E$_g$: 198 cm$^{-1}$, B$_1g$: 398
cm$^{-1}$, A$_1g$ and B$_1g$: 515 cm$^{-1}$, and E$_g$: 640
cm$^{-1}$. The rutile phase has four Raman active modes, the
predominant ones are E$_g$: 448 cm$^{-1}$, A$_1g$: 612 cm$^{-1}$,
and B$_1g$: 827 cm$^{-1}$. Since the Raman peaks from each phase
are clearly separated in frequency, the two phases are easily
distinguishable by their Raman spectra.

\begin{figure}
\unitlength1cm
\begin{picture}(5.0,6.)
\put(-3,-0.3){ \epsfig{file=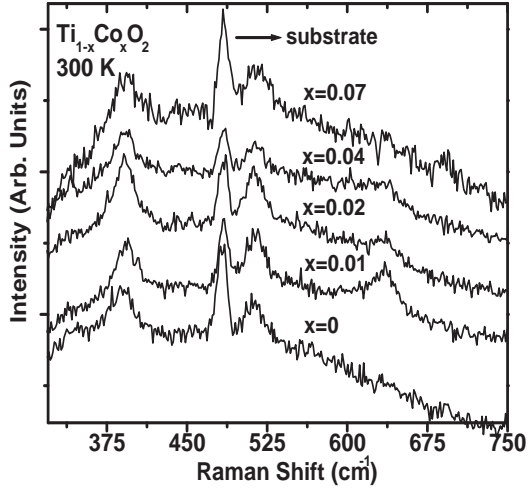, angle=0, width=8cm,
totalheight=8cm}}
\end{picture}
\caption{Raman spectra of Ti$_{1-x}$Co$_{x}$O$_2$ for values of
$x$ at 300 K. The peak at 485 cm$^{-1}$ is from the substrate.}
\label{fig2}
\end{figure}

Figure \ref{fig2} shows the Raman spectra from
Ti$_{1-x}$Co$_x$O$_2$ for various Co concentrations. The 485
cm$^{-1}$ peak is from the LaAlO$_3$ substrate and serves as a
good calibrant. Since the notch filter cuts off energy positions
below 200 cm$^{-1}$, we do not observe the low energy E$_g$ Raman
peaks. The Raman peaks at 398 cm$^{-1}$ and 515 cm $^{-1}$ clearly
show that the samples are in the anatase phase. A signature of the
640 cm$^{-1}$ is observed for $x$ =0.01 and 0.02 samples. Within
the experimental accuracy of a few wavenumbers there is no
discernable shift of the Raman peaks of the sample with higher
Co-doping concentration. However, for the higher Co-doped samples
the Raman peaks broaden. This may be an indication of limited Co
solubility resulting in a disorder of the crystal structure. A
recent work by Shinde \textit{et al.} shows that higher Co-doping
(beyond $x$ =0.02) without annealing results in the formation of
Co clusters in Ti$_{1-x}$Co$_x$O$_2$.\cite{shinde}

Though the absorption edge of anatase TiO$_2$ is at $\sim$ 3.2 eV,
the luminescence spectrum Stokes shifts by almost 1 eV. A broad
greenish-yellow emission, centered at 2.3 eV is observed from
films and crystals of anatase TiO$_2$.\cite{tang1,kiisk} The
TiO$_6$ octahedra in the anatase phase are distorted to lower
symmetry, which lifts degeneracies and creates band splitting
resulting in a narrowing of the conduction band. The band-to-band
excitation therefore results in self-trapped excitonic states due
to a localization of the excited state and its energy is lowered
by the lattice relaxation energy.\cite{tang1}

\begin{figure}
\unitlength1cm
\begin{picture}(5.0,6.)
\put(-3,-0.3){ \epsfig{file=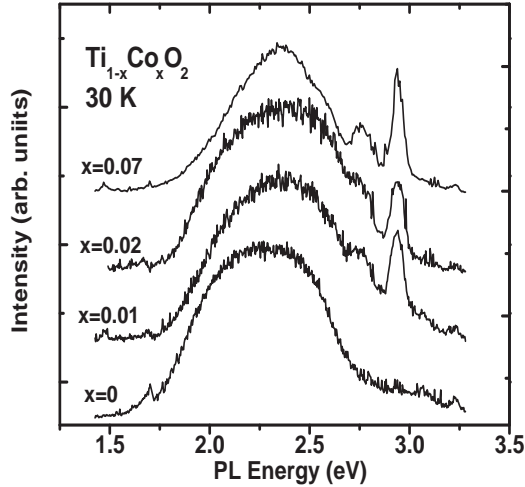, angle=0, width=8cm,
totalheight=8cm}}
\end{picture}
\caption{PL spectra of Ti$_{1-x}$Co$_{x}$O$_2$ for various Co
concentration at 30 K.} \label{fig3}
\end{figure}

Figure \ref{fig3} shows the PL spectra of Ti$_{1-x}$Co$_x$O$_2$
for various values of $x$, measured at 30 K. The spectra are
characterized by a broad emission centered at 2.3 eV due to the
self-trapped excitons. The individual spectra have been normalized
to the 2.3 eV emission and vertically shifted for clarity. A
recent work using polarized PL shows that this broad emission can
be further resolved into at least two peaks centered at 2.2 eV and
2.4 eV.\cite{kiisk} The center of the self-trapped emission (STE)
shows a slight blueshift in energy with increased doping
concentration.

The most striking feature of Fig. \ref{fig3} is the appearance of
two peaks at 2.77 eV and 2.94 eV in the Co-doped samples which are
absent in the undoped samples. At 30 K the 2.77 eV peak appears as
a weak shoulder in the low doped samples. To understand the origin
of the above peaks, we have carried out a thorough temperature
dependence of the PL. Since higher cobalt doping leads to
clustering in as-grown films,\cite{shinde} we compare the PL from
undoped TiO$_2$ to the low Co-doped samples as a function of
temperature.

\begin{figure}
\unitlength1cm
\begin{picture}(5.0,6.)
\put(-3,-0.3){ \epsfig{file=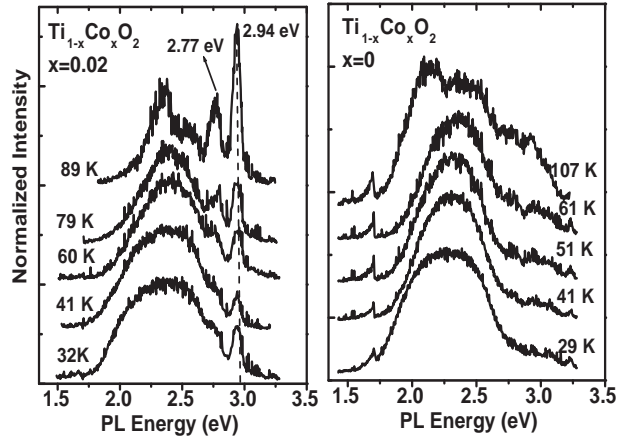, angle=0, width=10cm,
totalheight=8cm}}
\end{picture}
\caption{PL spectra of (a) Ti$_{0.98}$Co$_{0.02}$O$_2$ and (b)
undoped TiO$_2$ at selected values of temperature.} \label{fig4}
\end{figure}

Figure \ref{fig4} shows the PL spectra for a few selected values
of temperature for (a) $x$=0.02 and (b) $x$=0.0 samples. The
spectra have been normalized to the STE emission for both (a) and
(b). With increasing temperature the intensity of the STE drops
and beyond 120 K it is barely visible. For the $x$=0.02 sample the
energy position of STE shows almost no shift with increasing
temperatures. This is in contrast to other doped system, such as
Al-doped anatase TiO$_2$, where the STE redshifts by more than 500
mev from 5 K to 150 K due to trapping of excitons at higher
temperatures by Al acceptor levels.\cite{tang1} Our results
clearly show that the cobalt impurities do not result in any
additional self-trapping mechanism.

The absolute intensities of the 2.77 eV and 2.94 eV PL peaks do
not change significantly with temperature. Since the PL spectra
are normalized to the STE peak in Fig. \ref{fig4} (a), it appears
that the 2.94 eV and the 2.77 eV peaks gain intensity upon
increasing temperatures. These two peaks are not related to any
oxygen vacancies. We have compared the PL spectra (at different
temperatures) from two halves of the $x$=0.02 sample: one half
being the as-grown and the other half was annealed in an oxygen
atmosphere. The PL spectra at all temperatures from both samples
are identical, eliminating any issues related to oxygen vacancies.
The difference in energy of 170 meV between the 2.77 eV and the
2.94 eV peaks may be related to states with a given spin and a
spin-flipped state.

Band structure calculations show that the valence band derives
primarily from oxygen p-levels, the conduction band from the Ti
d-levels and the crystal-field split Co d-levels fall within the
energy gap. The local-spin-density approximation (LSDA)
calculations by Park \textit{et al.} show that the crystal field
splitting between $t_{2g}$ and $e_{g}$ states is larger than the
exchange splitting between $t_{2g}$ states, suggesting a low spin
state of Co.\cite{park} The appearance of the 2.77 eV and the 2.94
eV energy levels in our PL spectra appear well below the
absorption edge of 3.2 eV in Ti$_{1-x}$Co$_x$O$_2$,\cite{simpson}
indicating that they most probably arise from the cobalt $e_{g}$
energy levels. Moreover, the 170 meV difference between the two PL
peaks agree well with the energy separation between the spin-up
and spin-down $e_{g}$ states of the LSDA calculation.\cite{park}

The location of Co midgap levels also depends on the nature of the
microscopic distribution of the co atoms, as seen in the work by
Yang \textit{et al.}\cite{yang} The local oxygen non-stoichiometry
(which influences the valence states of cations) also affects the
band filling. Thus, the assignment of the observed levels to
specific electronic states would require these inputs. Based on
the total energies calculated for various configurations, Yang
\textit{et al.} suggest that cobalt ions in the
Ti$_{1-x}$Co$_x$O$_2$ matrix would prefer non-uniform doping mode
with short Co-Co distances.\cite{yang} Future work involving
magneto-optical techniques should reveal the nature of magnetic
coupling and magnetic excitations from various
Ti$_{1-x}$Co$_x$O$_2$ configurations and their possible
participation in optical processes.


\begin{acknowledgments}
S.G. acknowledges the donors of the American Chemical Society
Petroleum Research Fund Grant No. 38193-B7 and the Research
Corporation Grant No. CC5332. Work at UMD is supported under
NSF-MRSEC DMR Grant No.00-80008.

\end{acknowledgments}



\end{document}